\documentclass[pra,aps,a4paper,twocolumn,repreprint, superscriptaddress, longbibliography]{revtex4-2}

\usepackage{amssymb}
\usepackage{amsmath}
\usepackage{amsfonts}
\usepackage{graphicx}
\usepackage{color, soul}
\usepackage{natbib}
\usepackage{hyperref}
\usepackage{adjustbox}
\usepackage{tabularx}
\usepackage{breqn}
\usepackage{multirow}

\begin{document}

	\title{Phonon heat capacity and self-heating normal domains in NbTiN nanostrips}

	\author{M. Sidorova} 
	\affiliation{Humboldt-Universität zu Berlin$,$ Department of Physics$,$ Newtonstr. 15$,$ 12489 Berlin$,$ Germany}
	\affiliation{German Aerospace Center (DLR)$,$ Institute of Optical Sensor Systems$,$ Rutherfordstr. 2$,$ 12489 Berlin$,$ Germany}
	
	\author{A.D. Semenov}
	\affiliation{German Aerospace Center (DLR)$,$ Institute of Optical Sensor Systems$,$ Rutherfordstr. 2$,$ 12489 Berlin$,$ Germany}
	
	\author{H.-W. Hübers}
	\affiliation{Humboldt-Universität zu Berlin$,$ Department of Physics$,$ Newtonstr. 15$,$ 12489 Berlin$,$ Germany}
	\affiliation{German Aerospace Center (DLR)$,$ Institute of Optical Sensor Systems$,$ Rutherfordstr. 2$,$ 12489 Berlin$,$ Germany}

	\author{S. Gyger}
	\affiliation{Department of Applied Physics$,$ KTH Royal Institute of Technology$,$ SE-106 91 Stockholm$,$ Sweden}
	
	\author{S. Steinhauer}
	\affiliation{Department of Applied Physics$,$ KTH Royal Institute of Technology$,$ SE-106 91 Stockholm$,$ Sweden}

	\begin{abstract}
	Self-heating normal domains in thin superconducting NbTiN nanostrips were characterized via steady-state hysteretic current-voltage characteristics measured at different substrate temperatures. The temperature dependence and the magnitude of the current, which sustains a domain in equilibrium at different voltages, can only be explained with a phonon heat capacity noticeably less than expected for 3-d Debye phonons. This reduced heat capacity coincides with the value obtained earlier from magnetoconductance and photoresponse studies of the same films. The rate of heat flow from electrons at a temperature $T_e$ to phonons in the substrate at a temperature $T_B$ is proportional to $(T_e^p - T_B^p)$ with the exponent $p \approx 3$, which differs from the exponents for heat flows mediated by the electron-phonon interaction or by escaping of 3-d Debye phonons via the film/substrate interface. We attribute both findings to the effect of the mean grain size on the phonon spectrum of thin granular NbTiN films. Our findings are significant for understanding the thermal transport in superconducting devices exploiting thin granular films.
	\end{abstract}

\date{\today}
\maketitle

\section{Introduction}

Low-dimensional superconducting structures (e.g., thin films, nanowires, nanotubes, nanoparticles, and superlattices) have become building blocks for various fascinating applications such as single-photon detectors, hot-electron bolometers, kinetic inductance detectors, quantum interference devices, quantum bits (qubits), and other circuit elements \cite{esmaeil2021superconducting, shurakov2015superconducting, baselmans2012kinetic, fagaly2006superconducting, clarke2008superconducting, devoret2013superconducting}. For the design and optimization of these elements, the knowledge of transport and thermodynamic properties is crucial while qualitative understanding of the impact of reduced dimensionality on these properties is of fundamental interest. Since, at sufficiently low temperatures, the phonon wavevectors perpendicular to the film plane become restricted by the reciprocal film thickness, the phonon spectrum undergoes strong modifications that manifest in a considerable change of the thermal conductivity \cite{li2003thermal} and heat capacity \cite{prasher1999non} of phonons. Although the size effect in the phonon heat capacity has been studied theoretically (for films \cite{prasher1999non}, nanowires \cite{zhang2007phonon}, and spherical grains \cite{baltes1973specific, lautehschlager1975improved}), the available models are limited to crystalline specimens. Most superconducting devices, however, utilize amorphous and polycrystalline, granular materials. 

We have recently reported \cite{sidorova2020electron, sidorova2021magnetoconductance} a reduction of phonon heat capacities in thin polycrystalline granular films at low temperatures that was attributed to the effect of the mean grain size. This result was derived from extensive studies of the magnetoconductance and photoresponse of thin NbTiN, NbN, and WSi films. To support these findings and to verify the consistency of techniques in the steady state and in the time domain, here we analyze hysteretic current-voltage characteristics (CVCs) of superconducting NbTiN strips. Hysteresis appears in the regime of current bias as the difference between the experimental critical (switching) current $I_C$ and the return current $I_r$, at which the strip returns to the superconducting state when the bias current in the normal state is gradually decreased. In the voltage-bias regime, $I_r$ defines the current plateau which is commonly affiliated to a self-heating normal domain with the length controlled by the applied voltage. The domain remains in equilibrium, which is set by the balance between Joule heating of electrons by $I_r$ and their cooling via heat diffusion along the strip and heat flow through its interfaces. 

The model of the self-heating normal domain describing hysteresis was first proposed in \cite{skocpol1974self}. For small differences between the superconducting transition temperature, $T_C$, and the bath (substrate) temperature, $T_B$, the authors admitted  that the heat flow from the strip to the underlying substrate is $Q \propto (T - T_B)$ where $T$ is the strip temperature. In order to cover larger differences between $T_C$ and $T_B$, this model was further modified by an introduction of the heat flow  $Q = K (T^p - T_B^p)$, where $K$ is the effective thermal conductance. The approach with $p=4$ was implemented in \cite{yamasaki1979self}. This value of the exponent $p$ is provided by a microscopic model \cite{little1959transport}  for the heat flow between two solids across their interface via 3-d Debye phonons. Further modifications of the self-heating normal domain model were made by incorporating the state (normal/superconducting) and temperature dependence of the electron thermal conductivity \cite{dharmadurai1979simplified, tinkham2003hysteretic, hazra2010hysteresis, maneval2012temperature} or by varying the exponent $p$ \cite{li2011retrapping, dane2021self}. Microscopic models \cite{little1959transport, kaganov1957relaxation, bezuglyi1997kinetics} describing the heat flow from electrons to phonons in the film and further to phonons in the substrate show that the exponent $p$ is not necessary an integer and may have any value from 4 to 6. The value of $p$ in particular film is controlled by the dimensionality of phonons and by the degree of disorder. 

In this study, we analyze hysteretic CVCs of granular disordered NbTiN strips with different thicknesses. We apply a modified model of the self-heating normal domain with an arbitrary exponent $p$ in order to account for the effect of disorder and the effect of the mean grain size on the heat flow. Furthermore, we considered the heat transfer between three different systems (electrons and phonons in the film and phonons in the substrate) and address the microscopic meaning of the exponent $p$ and the effective thermal conductance.

\section{Experiment and results}

\begin{figure}[b!]
	\centerline{\includegraphics[width=0.47\textwidth]{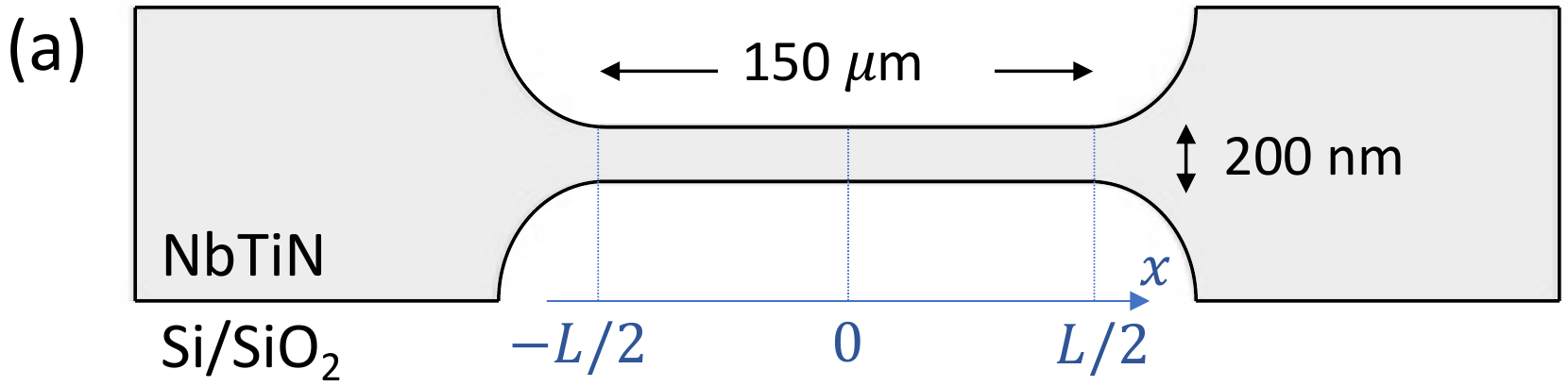}}
	
	\centerline{\includegraphics[width=0.5\textwidth]{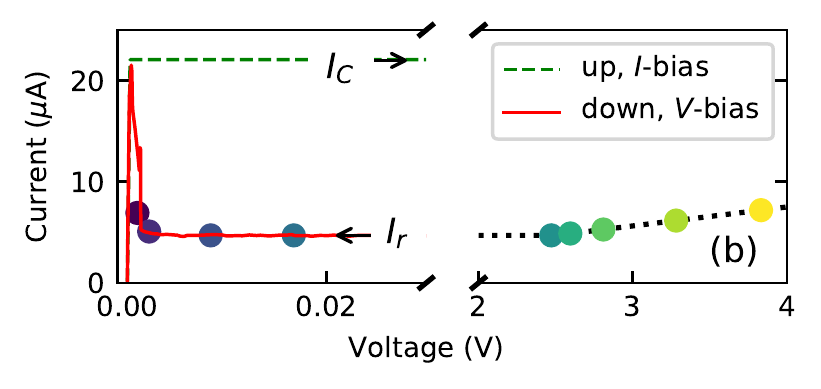}}
	\centerline{\includegraphics[width=0.5\textwidth]{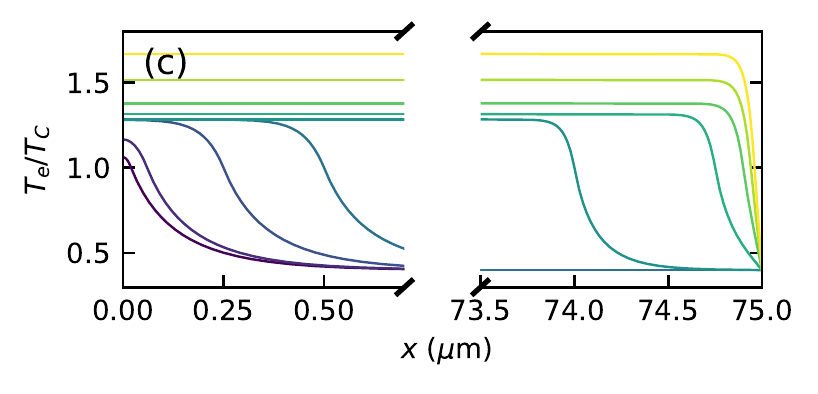}}
	\caption{(a) Sketch of the studied specimens (not in scale). (b) Curves (solid and dashed) represent CVCs of the NbTiN strip with $d=6$~nm measured at the bath temperature 2.9 K. The legend indicates the sweep directions and the corresponding bias regimes. Symbols represent discrete CVC points numerically computed (Eqs.~(\ref{eq:norm_dom}), boundary conditions (i-iii)) with actual strip parameters, $\lambda_S = \lambda_N(T_C)$ and the best-fit value $p=3.2$. The dotted curve is to guide the eyes. (c) Profiles of the electron temperature along the strip numerically computed for the discrete CVC points shown in panel (b); $x=0$ corresponds to the center of the strip and of the the normal domain. Curve colors in panel (c) correspond to the symbol colors in panel (b).} 
	\label{fig:IV_coord}
\end{figure}

The strips were fabricated from two NbTiN films with thicknesses 6 and 9~nm, which were studied earlier in \cite{sidorova2021magnetoconductance}. The films were deposited on Si substrates on top of a 270 nm-thick thermally-grown SiO$_2$ layer. They were shaped into straight strips with a length, $L$, of 150~$\mu$m and a width, $w$, of 200~nm. In order to reduce current crowding, the strips were terminated by tapered contacts. The shape of our specimens is shown in Fig.~\ref{fig:IV_coord}(a). Details of the fabrication procedure have been reported elsewhere \cite{steinhauer2020nbtin}. The specimens were mounted in a closed compartment inside a continuous-flow cryostat. The steady-state CVCs were measured at a set of fixed bath temperatures, $T_B$, between $2.5 - 8$~K.

Fig.~\ref{fig:IV_coord}(b) shows typical CVCs for our strips measured in two regimes: sweeping current from zero value upwards (current-bias) and sweeping voltage from a value in the resistive state downwards (voltage-bias). The former regime reduces the impact of bias electronics on the switching current $I_C$ while the latter reveals the current plateau at the return current $I_r$ caused by the presence of an equilibrium normal domain in the strip.
Adopting the heat flow from electrons to the substrate in the form $Q_\epsilon = K(T_e^p - T_B^p)$ ($T_e$ is the electron temperature, $K$ is the effective thermal conductance, see the next section) and assigning a constant electron thermal conductivity $\lambda = D\,c_e(T_C)$ ($D$ is the electron diffusivity, $c_e$ is the electron heat capacity, Table~\ref{tab:NbTiN_parameters}) to the normal and superconducting parts of the strip, we solved numerically the system of steady-state heat balance equations (Eqs.~(\ref{eq:norm_dom})). The obtained $T_e(x)$ profiles along the strip (Fig.~\ref{fig:IV_coord}(c)) were computed with the best-fit value $p=3.2$ (see below) for a set of lengths of the normal domain with the edges at $T_e = T_C$. Knowing the length and, hence the resistance of the domain, we further obtained the voltage along the strip. Corresponding discrete CVC points are shown with symbols in Fig.~\ref{fig:IV_coord}(b).

\newcolumntype{c}{>{\centering\arraybackslash}X}
\begin{table*}[t!]
	\centering
	\caption{Parameters of the NbTiN strips studied here. Electron and phonon heat capacities and the electron-phonon energy relaxation times at the transition temperatures ($T_C$) of the strips were obtained via extrapolation of corresponding values of non-structured films \cite{sidorova2021magnetoconductance} according to $c_e \propto T$ (Drude model), $c_{ph} \propto T^3$ (Debye model) and $\tau_{EP} \propto T^{-n}$, respectively. The values of the exponent $n$ were reported in \cite{sidorova2021magnetoconductance}. $R_\square$ is the normal-state sheet resistance, $D$ is the electron diffusivity, $\tau_{esc}$ is the phonon escape time, and $\tau_\epsilon$ is the relaxation time of the electron energy (see Appendix~\ref{app:tau_en}).}
	\label{tab:NbTiN_parameters}
	\begin{tabularx}{\textwidth}{@{}cccccccccccc@{}}
		\hline \hline
		Strip & $d$	&$w$ & $T_C$ &$R_\square$  & $D$   & $\tau_{esc}$ & $\tau_{EP}(T_C)$  &n& $c_e(T_C)$  & $c_{ph}(T_C)$& $\tau_\epsilon(T_C)$ \\
		& (nm)  & (nm) & (K) & ($\Omega$/sq) & (cm$^2$/s) & (ps)&   (ps)& & (J/Km$^3$) &  (J/Km$^3$) & (ps)  \\ 
		\hline
		NbTiN&6&200& 7.75   &710.6 &0.458& 53  & 6.5&3.5&  970 & 2115& 78\\
		NbTiN&9&200& 8.60&381.4 &  0.472& 80  & 4.0& 3.4& 1303& 8173& 90\\
		\hline \hline
	\end{tabularx}
\end{table*}

\begin{figure}[b!]
	\centerline{\includegraphics[width=0.5\textwidth]{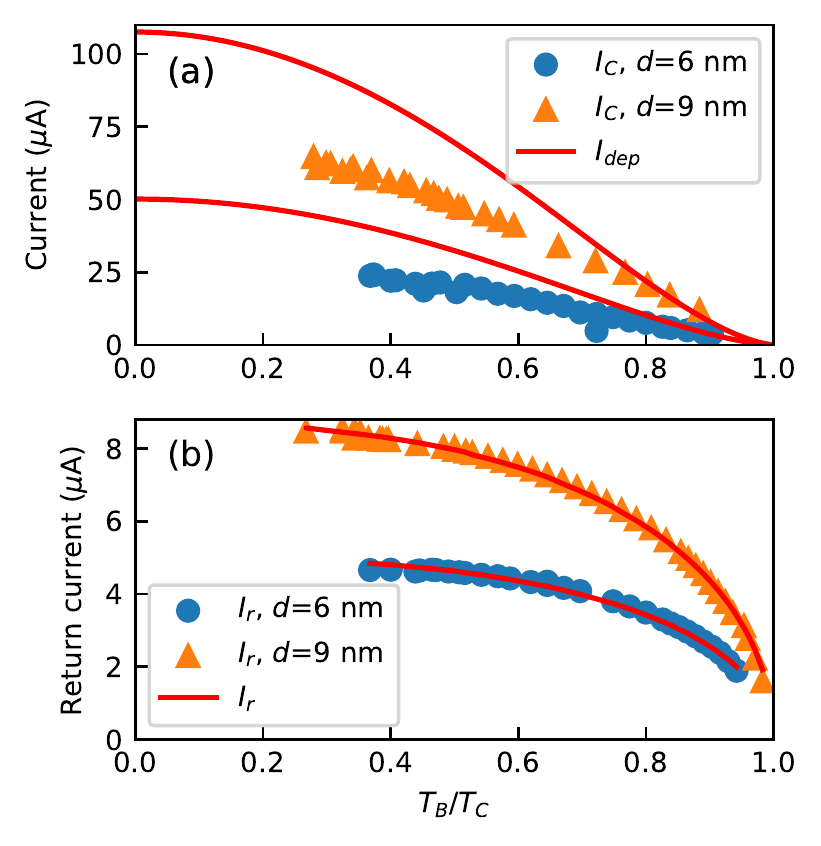}}
	\caption{Critical (a) and return (b) currents vs. fixed bath temperatures for two NbTiN strips. Symbols: experimental critical and return currents; solid curves: the best-fit theoretical depairing currents and return currents computed with Eq.~(\ref{eq:current_GL}) and Eq.~(\ref{eq:current}), respectively.}
	\label{fig:I_C_and_I_r}
\end{figure}

From experimental CVCs, we extracted $I_C(T_B)$ and $I_r(T_B)$ dependences, which are plotted in Fig.~\ref{fig:I_C_and_I_r}(a) and Fig.~\ref{fig:I_C_and_I_r}(b), respectively. For each strip, we computed the theoretical depairing current, $I_{dep}$, with Eq.~(\ref{eq:current_GL}) using the material parameters from Table~\ref{tab:NbTiN_parameters}. We fitted the $I_{dep}(T)$ dependences to the experimental $I_C(T_B)$ data in the vicinity of the superconducting transition (Fig.~\ref{fig:I_C_and_I_r}(a)) where both currents are expected to be equal. The value of $T_C$ was used as a fitting parameter in order to account for an expected reduction of the transition temperature in nanostructures as compared to non-structured films \cite{charaev2017proximity}. The best-fit values of $T_C$ listed in Table~\ref{tab:NbTiN_parameters} are indeed slightly smaller than those of non-structured films (8.41 and 9.51~K for films with thicknesses 6 and 9~nm, respectively \cite{sidorova2021magnetoconductance}). At the bath temperature of 3 K, the ratios $I_C / I_{dep}$ for our strips are 0.56 ($d=6$~nm) and 0.63 ($d=9$~nm). These values are comparable to those for other disordered thin films \cite{sidorova2018timing}.
From the combined analytical and numerical solution of the heat balance equations for the electron temperature, we computed $I_r(T_B)$ dependences (Eq.~(\ref{eq:current})) using the best-fit values of $T_C$ and the parameters from Table~\ref{tab:NbTiN_parameters}. The exponent $p$ and the scaling factor $A$ in Eq.~(\ref{eq:current})  were used as the only fitting parameters. The results shown in Fig.~\ref{fig:I_C_and_I_r}(b) with solid lines were obtained with $p=$ 3.2 and 2.9 for strip thicknesses 6 and 9~nm, respectively, and with $A\approx 1$ for both strips. 

\section{Discussion}

\begin{figure}[h!]
	\centerline{\includegraphics[width=0.5\textwidth]{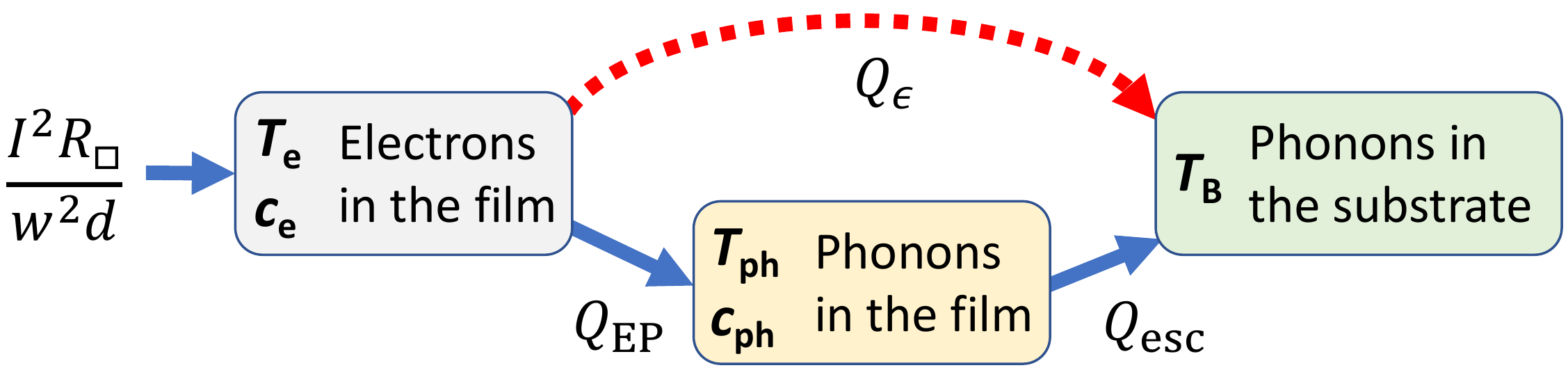}}
	\caption{Steady-state heat-flow diagram. $I^2 R_\square/(w^2 d)$ is the Joule power dissipated by the bias current $I$ per unit volume of the electron system; $Q_\epsilon$, $Q_{EP}$, and $Q_{esc}$ denote the net heat fluxes between different subsystems.}
	\label{fig:energy_diagram}
\end{figure}

The steady-state heat flow from a system at a temperature $T$ to a thermal bath with a temperature $T_B$  via a thermal link  is often described as $Q = K (T^p - T_B^p)$ where $K$ is an effective thermal conductance \cite{little1959transport, kaganov1957relaxation, bezuglyi1997kinetics, yamasaki1979self,skocpol1974self, dharmadurai1979simplified, tinkham2003hysteretic, hazra2010hysteresis, maneval2012temperature, li2011retrapping, dane2021self, elo2017thermal, baeva2021thermal}. For a small change in the system temperature  $\Delta T \ll T$, the rate of change in the heat flow is $dQ/dt \approx K p T^{p-1} dT/dt$. On the other hand, for a small deviation from the equilibrium, the relaxation is exponential, i.e. $dQ/dt = c\: dT/dt = c\: \Delta T/\tau$ where $\tau$ is the relaxation time and $c$ is the heat capacitance of the system. Hence, the effective thermal conductance is $K = c(T)/(p\:T^{p-1}\tau(T))$. It may depend on the system temperature by virtue of arbitrary temperature dependences $c(T)$ and $\tau(T)$. When applying this approach to a thin metal film on a dielectric substrate, one has to consider two subsystems, i.e., electrons and phonons in the film. Electrons are heated by the current with the rate $I^2R_\square/(w^2 d)$ per unit volume. They further transfer the energy to phonons, which in turn release it to the substrate as it is schematically depicted in Fig.~\ref{fig:energy_diagram}. There are two limiting cases \cite{bezuglyi1997kinetics} separated by the value of the ratio between the phonon-electron energy relaxation time $\tau_{PE}=\tau_{EP}c_{ph}/c_e$ and the phonon escape time $\tau_{esc}=4d/(u_s\bar{\alpha})$, where $\bar{\alpha}$ is the angle-averaged transmission of the film/substrate interface given by the acoustic mismatch model \cite{kaplan1979acoustic}, $u_s$ is the sound velocity, $\tau_{EP}$ is the electron-phonon energy relaxation time, and $c_{ph}$ and $c_e$ are the phonon and electron heat capacities, respectively. 

(i) In the limiting case of a thick film, $\tau_{esc} \gg \tau_{PE}$, reabsorption by electrons thermalizes nonequilibrium phonons at a temperature $T_{ph}$ which is slightly larger but close to the electron temperature $T_e$. Since in the steady-state conditions the net heat flows from electrons to phonons, $Q_{EP}$, and from phonons to the substrate, $Q_{ecs}$, are equal, the rate of heat removal from electrons per unit volume of the film can be represented as  either of these two fluxes. For crystalline metallic films, the net heat flux per unit volume from electrons to phonons in the film was described in \cite{kaganov1957relaxation} as $Q_{EP} = c_e/(5T_{e}^4\tau_{EP}) (T_e^5 - T_{ph}^5)$. The net heat flux from the film to the substrate via 3-d Debye phonons was described in \cite{little1959transport} as $Q_{esc} = c_{ph}(T_{ph})/(4T_{ph}^3\tau_{esc}) (T_{ph}^4 - T_{B}^4)$. Alternatively, for thick films with $T_e \approx T_{ph}$, the rate of the energy removal from electrons can be described as a one-stage process (Fig.~\ref{fig:energy_diagram}) as $Q_\epsilon \approx c_{ph}(T_e)/(p\:T_e^{p-1}\tau_{esc})\;(T_e^p - T_B^p)$. For 3-d Debye phonons in the film and in the substrate $p =$ 4 \cite{little1959transport}. This approach has been implemented in Ref.\cite{yamasaki1979self}.

(ii) In the opposite limiting case of a thin film, $\tau_{esc} \ll \tau_{PE}$, nonequilibrium phonons escape to the substrate without being thermalized that leads to overheating of electrons with respect to equilibrium phonons. 
Microscopic analysis of this essentially two-stage process has shown \cite{bezuglyi1997kinetics}, that even in this case the heat flow from electtrons to the substrate can be phenomenologically described as a one-stage process in the general form $Q_\epsilon = K (T_e^p - T_B^p)$ with $K = c_{e}(T_e)/(p T_e^{p-1}\tau_\epsilon(T_e))$. Here, $\tau_\epsilon$ is the electron energy relaxation time, which appears as the response time in photoresponse measurements, and $p = 5 + \gamma$ where $\gamma$ varies depending on the phonon dimensionality and the degree of disorder of the strip material. For 3-d Debye phonons and strongly disordered films, $\gamma =\,\,$1. We shall note here that the microscopic expression suggested in \cite{bezuglyi1997kinetics} for the effective thermal conductance $K$ is temperature independent. This fact imposes a constraint on the temperature-dependent quantities entering $K$: $c_{e}(T_e)$,  $\tau_\epsilon(T_e)$, and $T_e^{p-1}$. 

Although our NbTiN films with $\tau_{esc}/\tau_{PE} \approx 3.5$ fall into the intermediate regime between two limiting cases discussed above, we assume that even in this regime the heat flow from electrons to phonons in the substrate can be described as a one-stage process. Successful description of our experimental results validates the correctness of this assumption.   We apply the modified heat balance model to describe self-heating normal domain in a superconducting strip sustained in equilibrium due to the balance between Joule heating via the current $I_r$ and the cooling via thermal diffusion and heat flow through the film-substrate interface. The electron temperature distribution $T_e(x)$ along the strip is given by the solution of two steady-state heat-balance equations written for unit volumes of the normal (subscript $N$) and superconducting (subscript $S$) parts of the strip 
\begin{gather}
 - \frac{\partial}{\partial x}\left(\lambda_N\frac{\partial T_e}{\partial x}\right) + K_N \left(T_e^p - T_B^p\right) = \frac{I_r^2 R_\square}{w^2d},\,  |x|<x_{ND}\nonumber \\ 
 - \frac{\partial}{\partial x}\left(\lambda_S\frac{\partial T_e}{\partial x}\right) + K_S \left(T_e^p - T_B^p\right) = 0, \,  |x|>x_{ND}.
 \label{eq:norm_dom}
\end{gather}
The strip itself and its normal part (normal domain) are symmetrically centered at $x =0$. The coordinates of the edges of the normal domain $\pm x_{ND}$ are defined from the condition $T_e(x_{ND}) = T_C$. Both the electron thermal conductivities $\lambda_N$ and $\lambda_S$ as well as the effective thermal conductances $K_N$ and $K_S$ may be different in the normal and in the superconducting parts of the strip. Additionally, $\lambda$'s generally depend on temperature. 
The boundary conditions (BCs) for Eqs.~(\ref{eq:norm_dom}) are: (i) $(\partial T_e/ \partial x)_N = 0$ at  the center of the strip $x=0$; (ii) the temperature and the heat flux are continuous at the N/S interfaces where additionally the temperature equals the transition temperature, i.e.  $(T_e)_N = (T_e)_S = T_C$ and $\lambda_N(\partial T_e/ \partial x)_N = \lambda_S(\partial T_e/ \partial x)_S$ at $x=\pm x_{ND}$ ; (iii) $T_e = T_B$ at the ends of the strip $x=\pm L/2$. With these three BCs and $p\neq1$, the system Eqs.~(\ref{eq:norm_dom}) has no analytical solution.

In order to proceed analytically, we first assume that $\lambda$'s are temperature independent but different while the effective thermal conductances are the same in both parts of the strip $K_N = K_S = K$  and equal the value suggested in Ref.\cite{bezuglyi1997kinetics} for normal films. This simplifying assumption allows for a very accurate semi-analytical description of the experimental data. We discuss the impact of temperature dependent heat conductivity below.
With the simplifying assumptions, analytical solution for $p = 1$ \cite{skocpol1974self} and numerical for $p = 4$ \cite{yamasaki1979self} showed that for $ x_{ND}$, $(L/2-x_{ND})\gg L_T $, where $L_T$ is the effective thermal length defining the width of the domain edge, one can introduce two additional approximate BCs: (iv) $\partial T_e / \partial x = 0$ at the ends of the strip $x=\pm L/2$; and (v) $\partial ^2T_e / \partial x^2  = 0 $ at the strip center $x=0$. These additional BC's are easy to recognize in numerically computed and plotted in Fig.~\ref{fig:IV_coord}(c) $T_e(x)$ profiles for a set of different values of $x_{ND}$. They were obtained for $p = 3.2$ under the assumption $\lambda_S = \lambda_N = D\,c_e(T_C)$. In this case $L_T = \sqrt{D\,c_e(T_C)/K\,T_C^{p-1}}$. We also confirmed the validity of these additional BC's for temperature dependent $\lambda$'s (see the discussion below) and $p = 1$ by solving numerically Eqs.~(\ref{eq:norm_dom}).

For temperature independent but different $\lambda$'s, the first terms in Eqs.~(\ref{eq:norm_dom}) reduce to $\lambda_{N,S}\partial ^2T_e / \partial x^2$. Defining the Joule temperature as $T_J^p = I_r^2R_\square/(K w^2 d)$ and applying BC (v) to the first equation in Eqs.~(\ref{eq:norm_dom}), one gets
\begin{equation}\label{eq:T_B_T_0} 
T_{e0}^p = T_B^p + T_J^p,
\end{equation}
where $T_{e0} = T_e(x = 0)$.
 
One can further reduce Eqs.~(\ref{eq:norm_dom}) to two first-order differential equations via the substitution $(\partial T/\partial x)^2 = 2 \int dT \, (\partial^2 T / \partial x^2) $. Further applying remaining BC's (i) - (iv) to these first order equations and using Eq.~(\ref{eq:T_B_T_0}) one can analytically relate $T_J$ with $T_B$ as follows 
\begin{eqnarray}\label{eq:t_J_t_B} 
\frac{\lambda_S}{\lambda_N}T_B^{p+1} + \dfrac{p+1}{p} T_J^p T_C - (T_B^p + T_J^p)^{(p+1)/p} = 
\nonumber\\
= \left(1-\frac{\lambda_S}{\lambda_N}\right) \left(\frac{1}{p}T_C^{p+1} - \frac{p+1}{p}T_B^p T_C\right). 
\end{eqnarray}

Solving Eq.~(\ref{eq:t_J_t_B}) numerically and substituting $K = c_{e}(T_{e0})/(p T_{e0}^{p-1}\tau_\epsilon(T_{e0}))$ in the definition of the Joule temperature one obtains $T_J(T_B)$ and the return current as 
\begin{equation}\label{eq:current}
 I_r = A\:\sqrt{\frac{c_e(T_{e0}) w^2 d }{p\:\tau_\epsilon(T_{e0}) R_\square  T_{e0}^{p - 1}}T_J^p(T_B)}.
\end{equation}
This final expression contains an additional scaling factor $A$ which was used as one of the fitting parameters. We found (see the discussion below) that the maximum envisaged difference in $\lambda$'s relevant to the N/S interface does not noticeably affect the best fit value of the exponent $p$. We therefore used $\lambda_N = \lambda_S = \lambda_N(T_C)$ in Eqs.~(\ref{eq:norm_dom}) and Eq.~(\ref{eq:t_J_t_B}) to compute model $I_r(T_B)$ curves (Fig.~\ref{fig:I_C_and_I_r}(b)) and the profiles of the electron temperature (Fig.~\ref{fig:IV_coord}(c)). Other parameters were taken from Table~\ref{tab:NbTiN_parameters}. The $\tau_\epsilon$  was computed (Appendix~\ref{app:tau_en}) as a function of the electron-phonon energy relaxation time $\tau_{EP}$, the phonon escape time $\tau_{esc}$, and the ratio between electron and phonon heat capacities $c_e/c_{ph}$, which were taken from Table~\ref{tab:NbTiN_parameters}. For strips with thicknesses 6 and 9~nm, we obtained the best-fit values $p=3.2$ and 2.9  and $A = 0.91$ and 0.94, respectively. With these values of $p$ and with the Drude temperature dependence $c_e\propto T_e$, the constraint imposed by the temperature-independent effective thermal conductance $K$ \cite{bezuglyi1997kinetics} implies that $\tau_\epsilon\propto T_e^{-1}$. Such temperature dependence was indeed found at temperatures around $T_C$ for the $\tau_\epsilon$ computed in the framework of the two-temperature (2-T) model (Appendix~\ref{app:tau_en}). It is important to stress here that both the magnitude and the temperature dependence of $\tau_{\epsilon}$ around $T_C$ are noticeably affected by the magnitude of $c_{ph}$. We could satisfy the constraint on $K$ only by using $c_{ph}$ value found in \cite{sidorova2021magnetoconductance}, which is noticeably less than the Debye heat capacity expected for a given sound velocity. Our best fit values of the exponent $p$ require $\gamma\approx -2$ in the microscopic theory \cite{bezuglyi1997kinetics}. Recalling that for 3-d Debye phonons $\gamma = 1$, we attribute the change in $\gamma$ and the reduced phonon heat capacity to the size effect imposed by grains on the phonon spectrum in our thin granular NbTiN films. 

\begin{figure}[h!]
	\centerline{\includegraphics[width=0.5\textwidth]{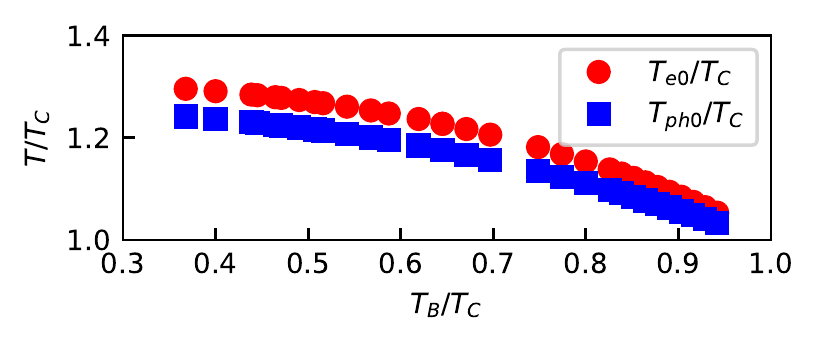}}
	\caption{Dimensionless electron and phonon temperatures in the center of the normal domain vs. dimensioneless bath temperature computed for the 6~nm-thick NbTiN strip. }
	\label{fig:temperatures}
\end{figure}

It's worth to emphasise here that for computing the return current with Eq.~(\ref{eq:current}), the heat capacities $c_e$ and $c_{ph}$ should be taken at the electron and phonon temperatures at $x=0$, $T_{e0}$ and $T_{ph0}$, and the energy relaxation time $\tau_\epsilon$ at $T_{e0}$. The phonon heat capacity, $c_{ph}$, enters Eq.~(\ref{eq:current}) through $\tau_\epsilon$, which is a function of $\tau_{EP}(T_{e0})$, $\tau_{esc}$, and $c_e(T_{e0})/c_{ph}(T_{ph0})$. In our computation procedure, $T_{ph0}$ was found from Eqs.~(\ref{eq:2TM}) in the steady-state conditions with $q=2+n$  where $n$ was taken from Table~\ref{tab:NbTiN_parameters} and $s=4$ (as for 3-d Debye phonons). Fig.~\ref{fig:temperatures} shows how the computed temperatures $T_{e0}$ and $T_{ph0}$ vary with the bath temperature.

We discuss now how temperature dependent $\lambda_N(T)$ and $\lambda_S(T)$ may affect the best fit value of the exponent $p$. In the normal state, $\lambda_N(T) = D c_e(T)\propto T$ by virtue of the $c_e(T)\propto T$ dependence. In the superconducting state, $\lambda_S(T)$ decreases much quicker. Down to the relative temperature $0.3 T_C$, with a good accuracy, it can be  approximated as $\lambda_S \approx \lambda_N \, 4/3 \,(T/T_C - 0.3)$ \cite{Abricosov}. Solving Eqs.~(\ref{eq:norm_dom}) numerically for $p=1$ with temperature dependent $\lambda$'s and with $\lambda_N = \lambda_S = \lambda_N(T_C)$, we found that introduction of the temperature dependent conductivities cause a change of the temperature in the center of the normal domain $T_{e0}$ but do not affect additional BCs (iv) and (v). We therefore used Eq.~(\ref{eq:t_J_t_B}) to evaluate the effect of different but temperature independent $\lambda_N$ and $\lambda_S$ for $p>0$. Setting the ratio $\lambda_S / \lambda_N = 0.5$, we found a 15$\%$ decrease in the best-fit value of the exponent $p$. Note that $\lambda_{N}$ and  $\lambda_{S}$ enter Eq.~(\ref{eq:t_J_t_B}) via the boundary condition at $x = x_{ND}$ where they are equal. Physically, the heat flow through this N/S interface can be affected by the temperature distribution in the layer with a thickness of the order of the electron-phonon thermal length $l_{EP} = \sqrt{D\tau_{EP}}$ which is about an order of magnitude less than the effective thermal length $L_T$. Since the temperature change around the domain edge is $T_{e0}-T_B$, the temperature difference in the layer with the thickness $l_{EP}$ is $\Delta T = l_{EP}/L_T(T_{e0}-T_B)\approx 1$~K. Corresponding change in $\lambda_S$ is less than $20\%$ which would cause a correction to the best-fit value of the exponent $p$ remaining beyond our experimental accuracy. We, therefore, neglected the temperature variations in the electron thermal conductivity and used $\lambda_S = \lambda_N = \lambda_N(T_C)$ for the description of our experimental data.

In the interpretation of photoresponse data with the 2-T model in \cite{sidorova2021magnetoconductance}, a possible contribution of diffusion cooling was neglected by the authors and $\tau_\epsilon$ was associated with the measured response time. In order to check the validity of this approximation, we numerically solved the time-dependent heat balance equation including diffusion for small temperature deviations and computed the response time as a function of the strip length. The dependence shown in Fig.~\ref{fig:tau_en}(b) (Appendix~\ref{app:tau_en}) supports the assumption that the diffusion cooling was negligible for strip lengths used in \cite{sidorova2021magnetoconductance}. 
 
\section{Conclusion}
We have analyzed the hysteretic current-voltage characteristics of straight nanostrips fabricated from thin granular NbTiN films. We have shown that the results can be quantitatively explained only with the phonon heat capacity $c_{ph}$, which is drastically reduced compared to the value expected for 3-d Debye phonons. The same reduced $c_{ph}$ was obtained earlier from photoresponse studies. This shows the compatibility of steady-state and time-resolving experimental approaches for the evaluation of heat capacities and spectra of phonons in nanostructured superconducting thin films. 

Furthermore, we have shown that the steady-state experimental approach is self-consistent since it yields the same temperature dependence for the relaxation time of the electron energy as the two-temperature model predicts. 

We have also observed that the heat flow from electrons to the substrate $\propto (T_e^p - T_B^p)$ with the exponent $p \approx 3$ which differs from both mediated by the electron-phonon interaction and by escaping of 3-d Debye phonons via film/substrate interface. 

This finding, along with the reduced $c_{ph}$, is attributed to the effect of the mean grain size on the phonon spectrum of thin granular films. Our results provide important insights into thermal transport in thin nanostrips exploited in superconducting devices and thus pave the way for improving their performance.

\section*{Acknowledgements:}

The authors greatly acknowledge the help of V. Zwiller in
the sample preparation.

\appendix

\section{Depairing current}

The Ginzburg-Landau depairing current with the dirty-limit correction of Kupryanov and Lukichev \cite{kupryanov1980temperature} $C(T)$ is given by \cite{semenov2015asymmetry}
\begin{gather}
I_{\text{dep}} = C(T) w \frac{4\sqrt{\pi}\exp(2\gamma)}{21\zeta(3)\sqrt{3}} \frac{\beta_0^2 (k_B T_C)^{3/2}}{e R_\square \sqrt{D\hbar}} \left[ 1 - \left( \frac{T}{T_C}\right)^2 \right]^{3/2} \nonumber\\
C(T) = 0.65 \left[ 3- \left( \frac{T}{T_C}\right)^5 \right]^{1/2},
\label{eq:current_GL}
\end{gather}
where $\gamma$~=~0.577, $\zeta(3)$~=~1.202, $e$ is the electron charge, $k_B$ is the Boltzmann constant and $\beta_0$ is the ratio between the energy gap and $k_B T_C$. Since this parameter for NbTiN is not known, we used the standard BCS value $\beta_0 =1.76$.

\section{Electron energy relaxation time}
\label{app:tau_en}
\subsection{Generalized two-temperature model}
\label{app:tau_en_2T}
 We find the relaxation time of the electron energy via electron-phonon interaction and phonon escaping to the substrate, $\tau_\epsilon$, using two-temperature (2-T) model generalized for arbitray exponents $q$ and $s$ and for large differences between $T_e$, $T_{ph}$ and $T_B$. Generalized time-dependent heat balance equations take the form
\begin{gather}
 		c_e(T_e)\dfrac{dT_e}{dt} = -\frac{c_e(T_e)}{qT_{e}^{q-1}\tau_{EP}} (T_e^q - T_{ph}^q) + P \nonumber\\
	c_{ph}(T_{ph})\dfrac{dT_{ph}}{dt} = \frac{c_e(T_e)}{qT_{e}^{q-1}\tau_{EP}} (T_e^q - T_{ph}^q)-\nonumber \\  - \frac{c_{ph}(T_{ph})}{sT_{ph}^{s-1}\tau_{esc}} (T_{ph}^s - T_{B}^s),
\label{eq:2TM}
\end{gather}
 where $P$ is the power dissipated per unit volume in the electron subsystem. Small periodic variations in the dissipated power $\Delta P e^{-j\omega t}$ ($\Delta P \ll P$ and $\omega$ is the circular frequency) cause small periodic oscillations of $T_e$ and $T_{ph}$, i.e. $\Delta T_e e^{-j\omega t}$ and $\Delta T_{ph} e^{-j\omega t}$. Substituting them in Eqs.~(\ref{eq:2TM}) and cancelling steady-state parts, one gets the linearized equations for $\Delta T_e$ and $\Delta T_{ph}$ identical to those suggested in \cite{perrin1983response}. The known solution of linearized equations for $\Delta T_e$  is given by \cite{perrin1983response},\cite{sidorova2020electron} 
\begin{equation}\label{eq:Te}
\Delta T_e(\omega) =  Re \left( \frac{\Delta P}{c_e}\frac{\tau_2 \tau_3}{\tau_1}
		\frac{(1 + j \omega \tau_1)}{(1 + j \omega \tau_2)(1 + j \omega \tau_3)} \right).
\end{equation}
The characteristic times are $\tau_1=(\Gamma_2+\Gamma_3)^{-1}$ and $\tau_{2,3}= \left[ \frac{1}{2} \sum_{i} \Gamma_i 
\left( 1  \mp \sqrt{1 - 4 \Gamma_1 \Gamma_3 / (\sum_{i} \Gamma_i)^2} \right) \right]^{-1}$. Here, $\Gamma_1 = \tau_{EP}^{-1}$, $\Gamma_2 = \Gamma_1 c_e/c_{ph} $, and $\Gamma_3 = \tau_{esc}^{-1}$. For an electron subsystem obeying only one relaxation time $\tau$, the solution would have the form $|\Delta T_e(\omega)|=1/\sqrt{1+(\omega\tau)^2}$. In this case,  $| \Delta T_e(1/\tau)|=|\Delta T_e(0)/\sqrt{2}|$. We use the very same criterion to define $\tau_\epsilon$ for $\Delta T_e(\omega)$ given by Eq.~(\ref{eq:Te}).

\begin{figure}[h!]
	\centerline{\includegraphics[width=0.5\textwidth]{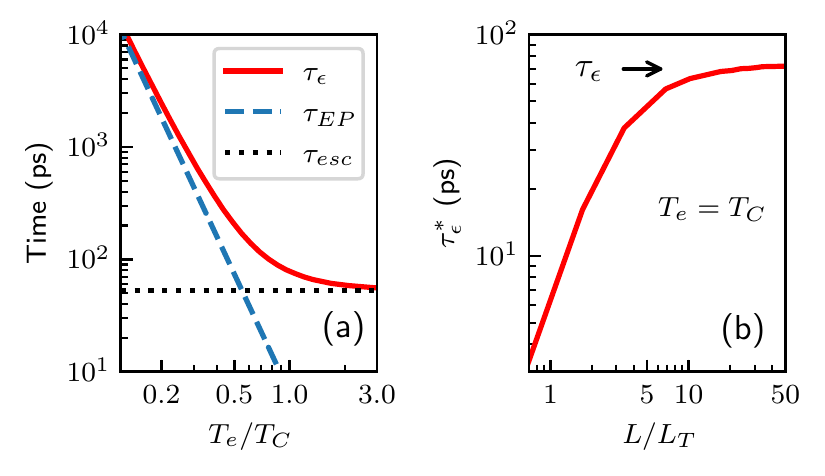}}
	\caption{ Electron energy relaxation time for 6~nm-thick NbTiN strip in the double-logarithmic scale. (a) $\tau_\epsilon$ vs. temperature computed with the uniform 2-T model without diffusion cooling. At low ($T_e \ll T_C$) and high ($T_e \gg T_C$) temperatures, $\tau_\epsilon$ asymptotically approaches $\tau_{EP}$ and $\tau_{esc}$, respectively. (b) $\tau_\epsilon^*$ vs. reduced strip length at $T_e = T_C$ computed with the heat balance equation including diffusion. For $L \gg L_T$, when diffusion cooling can be neglected, both models give the same relaxation time $\tau_\epsilon = \tau_\epsilon^*$.}
	\label{fig:tau_en}
\end{figure}

In Fig.~\ref{fig:tau_en}(a), we plot the temperature dependence of $\tau_\epsilon$ computed for the 6~nm-thick NbTiN film with $c_e(T)=c_e(T_C)(T_C/T)$ (Drude model), and $c_{ph}(T)=c_{ph}(T_C)(T_C/T)^3$ (Debye model), where $c_e(T_C)$ and $c_{ph}(T_C)$ are from Table~\ref{tab:NbTiN_parameters}. At small temperatures the electron energy relaxation time asymptotically approaches $\tau_{EP}\propto T^{-3.5}$ while at $T\gg T_C$ it saturates at the temperature independent value $\tau_{esc}$. Another important observation is that around $T_C$ we find $\tau_{EP}\propto T^{-1}$. This temperature dependence meets the constraint imposed by temperature-independent microscopic expression for the effective thermal conductance $K$.

\subsection{Impact of diffusion}
\label{app:tau_en_diff}

The 2-T model in the form of Eq.~(B1) does not account for the heat removal from electrons via electron diffusion, e.g., to the contacts. In order to check when the diffusion becomes important, we find the relaxation time of the total electron energy in the strip $\tau_\epsilon^*$ as a function of the strip length by numerically solving the linearized one-dimensional time-dependent heat balance equation with diffusion cooling $\tau_\epsilon \partial T_e/ \partial t = L_T^2 \partial^2 T_e/ \partial x^2  - (T_{e}-T_B)$, and applying a $\delta$-like uniform heat source and Dirichlet boundary conditions ($T_e = T_B$) at the strip edges. Here $L_T = \sqrt{D \tau_\epsilon}$ is the appropriate  ($p = 1$) thermal length. We further compute the weighted energy $E(t) = c_e\int\limits_0^{L/2} \delta T_{e}(t,x)\, dx$, where $\delta T_{e}(t,x)$ is the solution of the differential equation, and define $\tau_\epsilon^*$ from the condition $\delta E(\tau_\epsilon^*) = \delta E(0)/exp(1)$. As seen in Fig.~\ref{fig:tau_en}(b), diffusion affects $\tau_\epsilon^*$ when $L \leq 10\, L_T$. For the 6~nm and 9~nm-thick NbTiN films, $L_T \approx 60$~nm. The $\tau_\epsilon$ was obtained in \cite{sidorova2021magnetoconductance} for 1~$\mu$m-long NbTiN bridges, for which $L  > 10\, L_T$. Therefore, diffusive cooling had no impact on the energy relaxation time and, hence, on the phonon heat capacity evaluated in \cite{sidorova2021magnetoconductance}.

\bibliography{sample}

\end{document}